%% file: bare_jrnl.tex
\documentclass[journal]{IEEEtran}

\ifCLASSINFOpdf
\else
   \usepackage[dvips]{graphicx}
\fi
\usepackage{graphicx}
\usepackage{subfig}
\usepackage{booktabs}
\usepackage{amsmath}
\usepackage{amssymb}
\usepackage{url}
\usepackage{multirow}
\usepackage{balance}
\usepackage{tikz}
\usepackage{pgfplots}
\usepackage{pifont}

\newcommand{\cmark}{\ding{51}}
\newcommand{\xmark}{\ding{55}}

\begin{document}

\title{Inference-Adaptive Neural Steering for Real-Time Area-Based Sound Source Separation}

\author{Martin Strauss, Wolfgang Mack, Mar\'{i}a Luis Valero and Okan Köpüklü

\thanks{Martin Strauss is with the International Audio Laboratories (a joint institution of the Friedrich-Alexander-Universität Erlangen-Nürnberg and Fraunhofer IIS), 91058 Erlangen, Germany (e-mail: martin.strauss@audiolabs-erlangen.de), Wolfgang Mack is with Friedrich-Alexander-Universität Erlangen-Nürnberg email: wolfgang.mack@fau.de, Maria Luis Valero and Okan Köpüklü are with Microsoft Applied Sciences Group, Munich, Germany (email: okan.kopuklu@microsoft.com). Work done while the first author was doing an internship at Microsoft Applied Sciences Group.}}

\markboth{Journal of \LaTeX\ Class Files, Vol. 14, No. 8, August 2015}
{Shell \MakeLowercase{\textit{et al.}}: Bare Demo of IEEEtran.cls for IEEE Journals}
\maketitle

\begin{abstract}
We propose a novel \textit{Neural Steering} technique that adapts the target area of a spatial-aware multi-microphone sound source separation algorithm during inference without the necessity of retraining the deep neural network (DNN). To achieve this, we first train a DNN aiming to retain speech within a target region, defined by an angular span, while suppressing sound sources stemming from other directions. Afterward, a phase shift is applied to the microphone signals, allowing us to shift the center of the target area during inference at negligible additional cost in computational complexity. Further, we show that the proposed approach performs well in a wide variety of acoustic scenarios, including several speakers inside and outside the target area and additional noise. More precisely, the proposed approach performs on par with DNNs trained explicitly for the steered target area in terms of DNSMOS and SI-SDR.
\end{abstract}


\IEEEpeerreviewmaketitle

\section{Introduction}
\IEEEPARstart{T}{he} modern workplace is shifting from traditional office settings to open-plan offices, hybrid work models, and fully remote employment. For 2022, the German Federal Statistics Office reports that 24.2\% of employed people worked from home, with 14.7\% doing so at least half the time, nearly doubling from pre-pandemic levels~\cite{stat}. 

In this environment, virtual meetings are common, even in open offices, presenting challenges such as interfering speakers, background noise, and privacy concerns from recording non-participants.

A potential solution is to employ sound source separation techniques capable of separating the speech of meeting participants from all interfering sounds. Such an approach has two major requirements: (i) all speech utterances within a defined target spatial region should be kept to ensure all participants are covered 
and (ii) the employed solution should be real-time capable to prevent unwanted delays. Ideally, this target region is dynamically adaptable to other directions, allowing for the inclusion of more people and an intentional muting of individuals through steering away.

In this work, we present \textit{Neural Steering}, an inference-based adaption of a sound source separating deep neural network (DNN) to new target regions. This is achieved by applying a phase shift to the audio captured by a microphone array. For this purpose, we first train a DNN architecture that preserves all speech utterances within a region-of-interest (ROI) only defined by an angular span w.r.t. the origin/center of a linear microphone array and suppresses speakers outside the ROI and noises. Afterward, applying the proposed approach can shift the ROI with negligible computational cost while keeping the DNN's sound source separation ability.

The performance evaluation based on computational metrics shows that the proposed approach performs well in various settings, including multiple targets, interfering speakers, and noise. 

\section{Related work}

Source separation involves extracting individual audio sources from a mixture captured by one or more microphones.~\cite{source_sep}.
Multiple microphones enable the use of spatial filtering techniques, known as beamformers (BF), which utilize spatial information like time-difference-of-arrival~\cite{Benesty2008} to enhance signals from a specific direction while suppressing others.

Traditional BF methods~\cite{perrot2021so,gsc,lcmv,souden2009optimal,lotter2006dual} are well-established but their performance relies on accurate signal modeling or direction of arrival estimation~\cite{soumitro_2019,taseska_2017_multi_speaker_doa}.


With the advancements in DNNs, many combinations with BFs exist and enable non-linear spatial processing, e.g., used as a postfilter~\cite{tesch_2023,guided_se_2023} or standalone neural BFs~\cite{tesch_2024, Liu2022}.

Recently, there's been increasing interest in approaches that focus on the target signal's location, based on a predefined region~\cite{xu_2022, wechsler2023}, distance~\cite{yiwere_2019, patterson22_interspeech, Taherian2022}, or direction~\cite{Taherian2022, cos}. These methods are often impractical for real-time applications due to computational complexity or the assumption of a single target speaker with a known source direction. 


Our work builds on~\cite{strauss2024efficientareabasedspeakeragnosticsource} and aims to efficiently steer a pre-defined ROI to other directions, while being independent from the specific location and number of speakers. In contrast to e.g.~\cite{tesch_2024}, no source direction is needed. Our scenario is conceptually similar to~\cite{rezero}, but we use only a two-microphone setup and don't rely on sampling the target direction. 



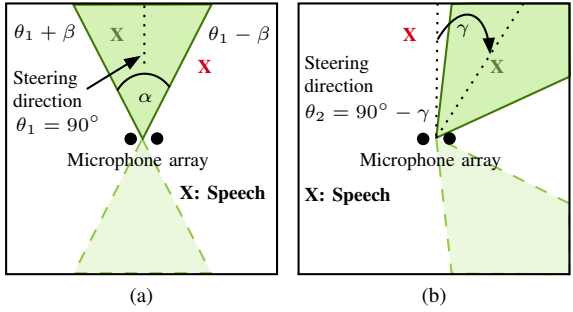
\begin{figure}[htb]
\centering
\makebox[0.4\textwidth][c]{%
\subfloat[]{\scalebox{1.05}{\input{setup.tikz}}}
\subfloat[]{\scalebox{1.05}{\input{setup_steered.tikz}}}}
\caption{Illustration of (a) the training setup for the area-based sound source separation and (b) applying Neural Steering to adapt the ROI to include the second speech source. Speech sources are denoted by $\textbf{X}$ (Green: target, red: interferer). The ROI is defined by the angle $\alpha$. The dashed lines denote the mirrored area due to front-back ambiguity of the ULA.}
\label{fig:setup}
\end{figure}

\section{Area-based sound source separation}
For area-based sound source separation, the ROI is spanned by an angle $\alpha$ in front of a microphone array with the center of the microphone array as origin. As the microphone array, we define a uniform linear array (ULA) with $M \in \mathbb{N}^+$ microphones placed at random location in a room.
Figure~\ref{fig:setup}\,(a) illustrates this setup for two microphones.

The target signal spectrum $\mathbf{X}_{t,m}(n,k)$ is determined by the summation of all speech signals inside the ROI in the short-time Fourier transform (STFT) domain, i.e.,
\begin{equation}
    \mathbf{X}_{t,m}(n,k) = \sum_{r=1}^R \mathbf{X}^{(r)}_{t,m}(n,k),
\end{equation}
\noindent where $r \in [1,...,R]$ is the speaker index and $n$, $k$ and $m$ denote the time frame, frequency bin and  microphone index. 

Respectively, all interfering sources $\mathbf{X}_{i,m}(n,k)$ are defined as the summed speech sources outside of the ROI, i.e.,
\begin{equation}
    \mathbf{X}_{i,m}(n,k) = \sum_{l=1}^L \mathbf{X}^{(l)}_{i,m}(n,k),
\end{equation}
\noindent with speaker index $l \in [1,...,L]$.

As a result, the STFT spectrum $\mathbf{Y}_m(n,k)$ of an audio mixture captured by the microphones is expressed as the summation
\begin{equation}
    \mathbf{Y}_m(n,k) = \mathbf{X}_{t,m}(n,k) + \mathbf{X}_{i,m}(n,k) + \mathbf{N}_m(n,k),
\end{equation}
\noindent including all target and interfering signals, in addition to non-speech-like noise $\mathbf{N}_m(n,k)$. The  noise can be located inside or outside the ROI.

The goal is to develop a DNN that learns to cover and extract signals from all active speech sources within the ROI. Therefore, it produces a separation mask $Q \in \mathbb{C}^{N\times K\times 1}$ to separate the estimated target spectrum $\hat{\textbf{X}}_t(n,k)$ from the input signal by an element-wise complex multiplication, denoted by $\odot$, to the reference microphone signal $\mathbf{Y}_{m=1}$, i.e,
    \begin{equation}
        \hat{\mathbf{X}}_t(n,k) = Q \odot \mathbf{Y}_{m=1}.
    \end{equation}

For this study we fix the number of microphones to be $M=2$ with the left microphone representing the reference.

\begin{figure}[htb]
\centering
\makebox[0.4\textwidth][c]{%
\subfloat[]{\scalebox{0.9}{\input{doa_2.tikz}}}\phantom{test}
\subfloat[]{\scalebox{0.6}{\input{arccos.tikz}}}}
\caption{In (a), a plane sound wave reaches the microphone array (incident angle $\theta$). The dashed line symbolizes the phase shift caused by the extra travel distance $l$ to the left microphone. (b) illustrates the $\arccos$ function to show the new ROI boundaries defined by the angles $\phi_{l,r}$ after shifting the steering vector.}
\label{fig:doa}
\end{figure}
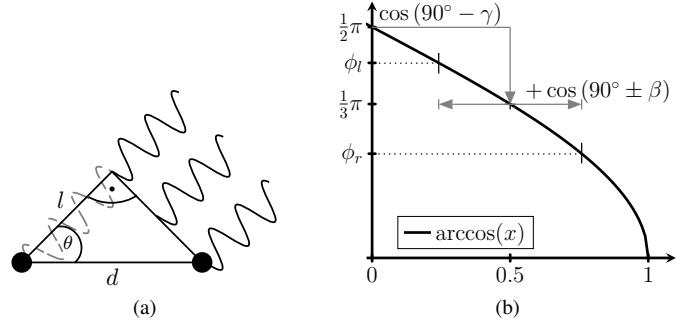

\section{Proposed Inference-Adaptive Neural Steering}
Steering to new target areas typically requires multiple DNNs for different ROIs, which is computationally costly. Instead, we propose training for a single ROI and shifting the entire ROI towards a new direction by applying a phase shift at inference time to the STFT spectrum of the second microphone. We term this approach  \textit{Neural Steering}, shown in Figure~\ref{fig:setup}\,(b). 

To obtain the phase shift, we consider a far- and free-field scenario. The wave originating from a sound source reaches the microphones at different times \cite{source_sep}. Let $d$ denote the distance between the two microphones and $\theta$ the incident angle, then the distance $l=d\cos(\theta)$ describes the additional way the incoming wave has to travel to the left microphone (see Figure~\ref{fig:doa}\,(a)). This results in the inter-microphone phase difference $\Delta\varphi_{\theta} = 2\pi f d \cos(\theta)/c$, where $f$ denotes the frequency and $c$ the speed of sound.

\begin{figure*}[!t]
\centering
  \makebox[0.15\textwidth][c]{%
  \subfloat[]{\scalebox{0.21}{\includegraphics{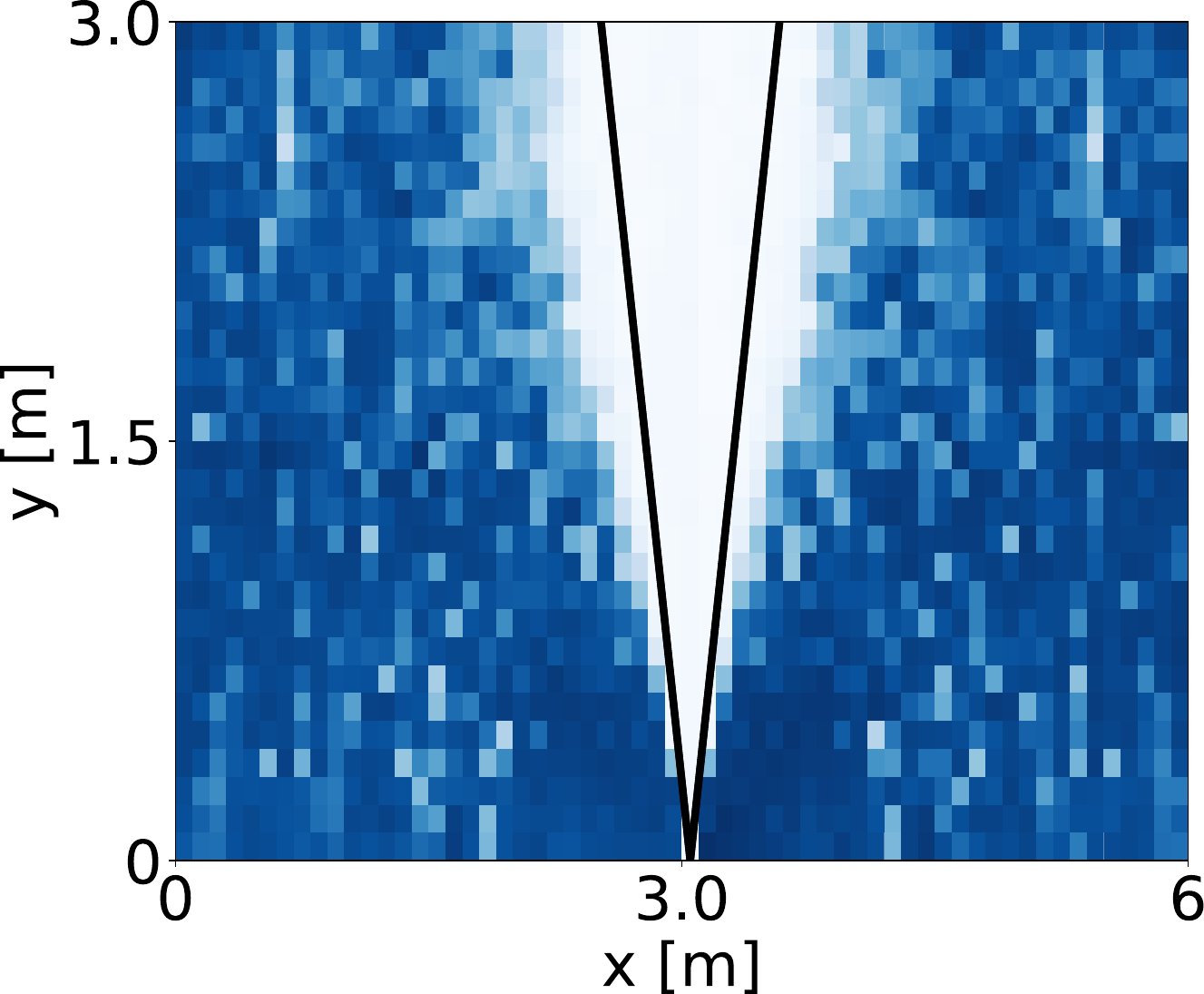}}}
  \hspace{1ex}
  \subfloat[]{\scalebox{0.21}{\includegraphics{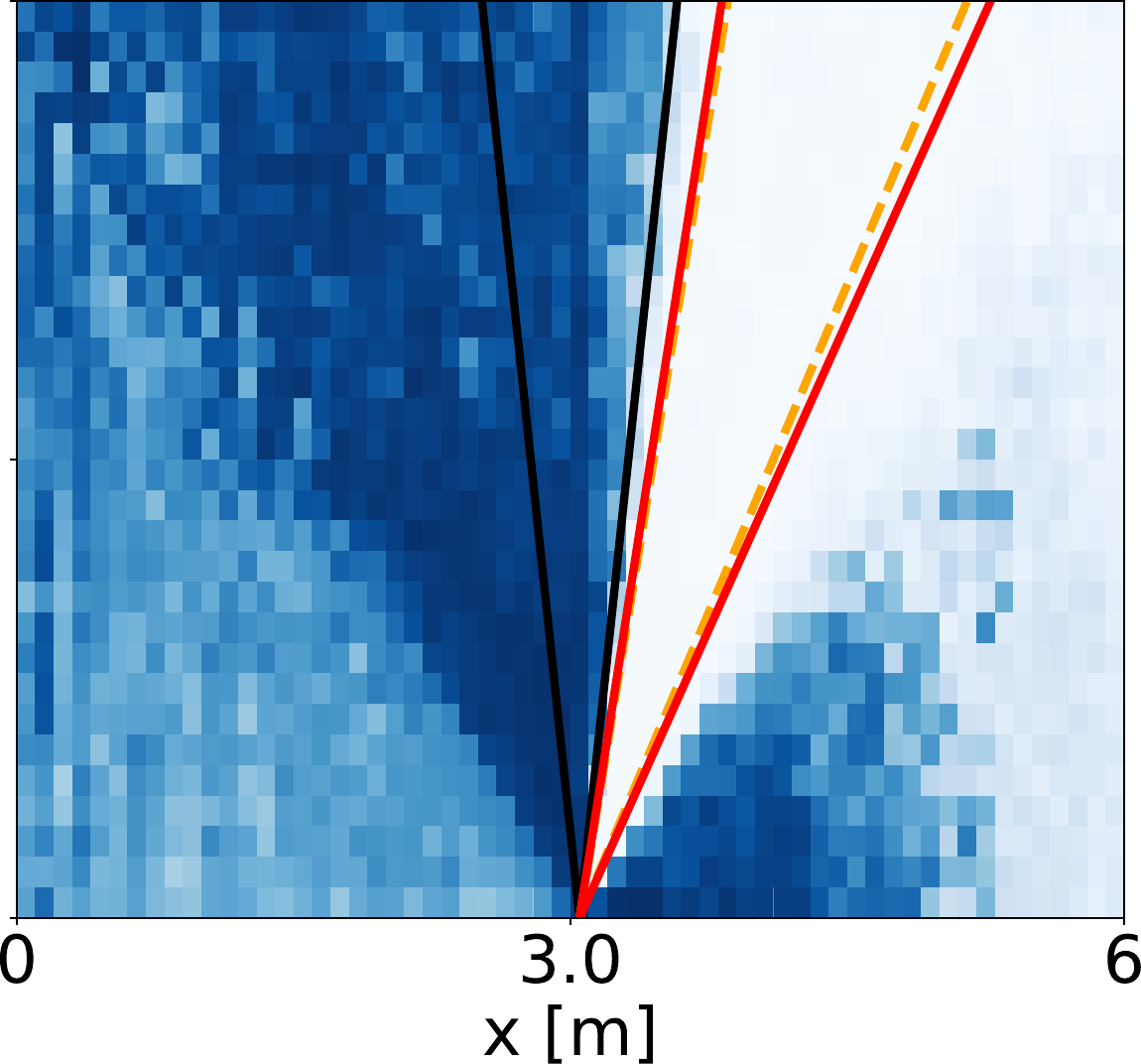}}}
  \hspace{1ex}
    \subfloat[]{\scalebox{0.21}{\includegraphics{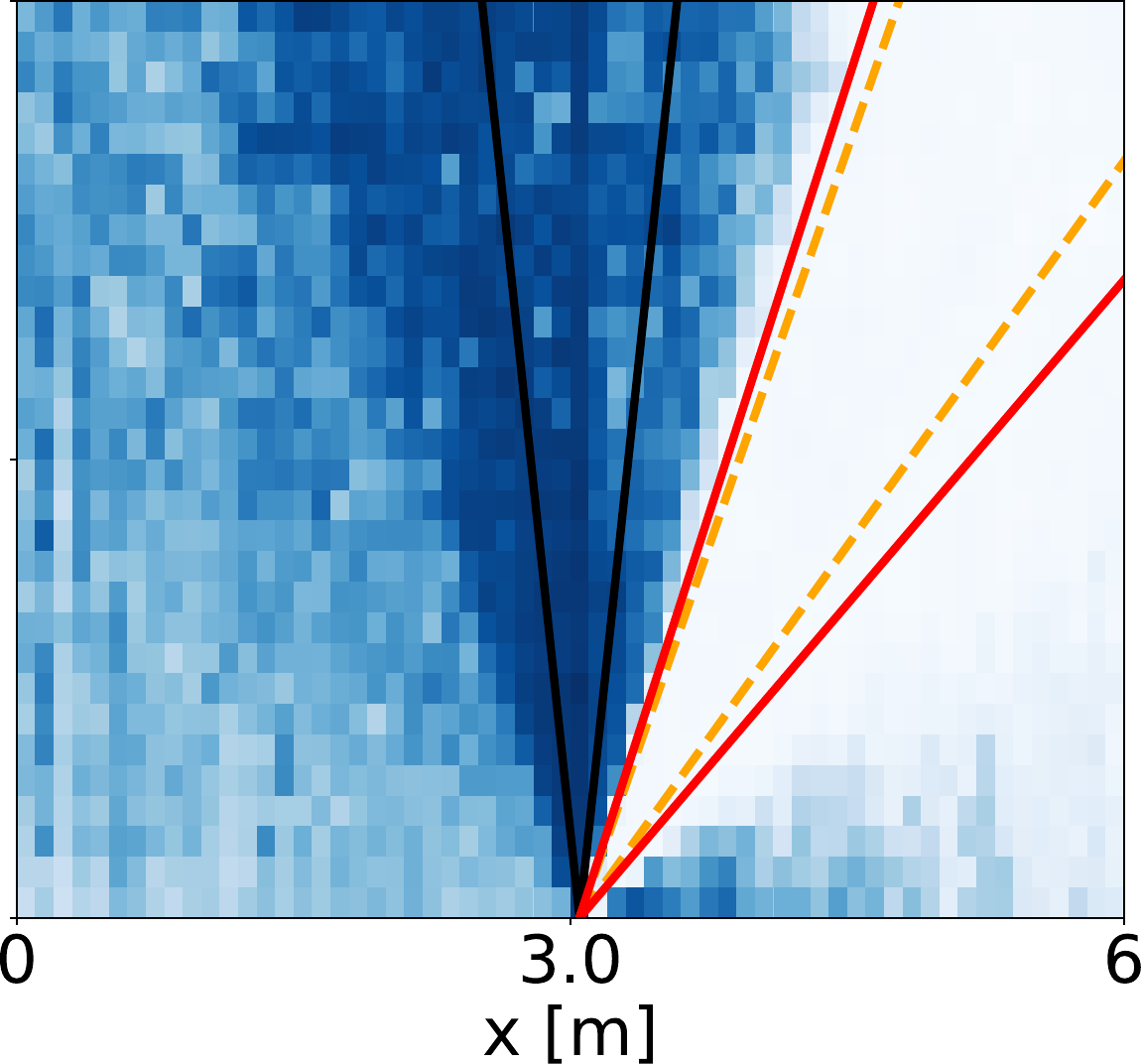}}}
    \hspace{1ex}
    \subfloat[]{\scalebox{0.21}{\includegraphics{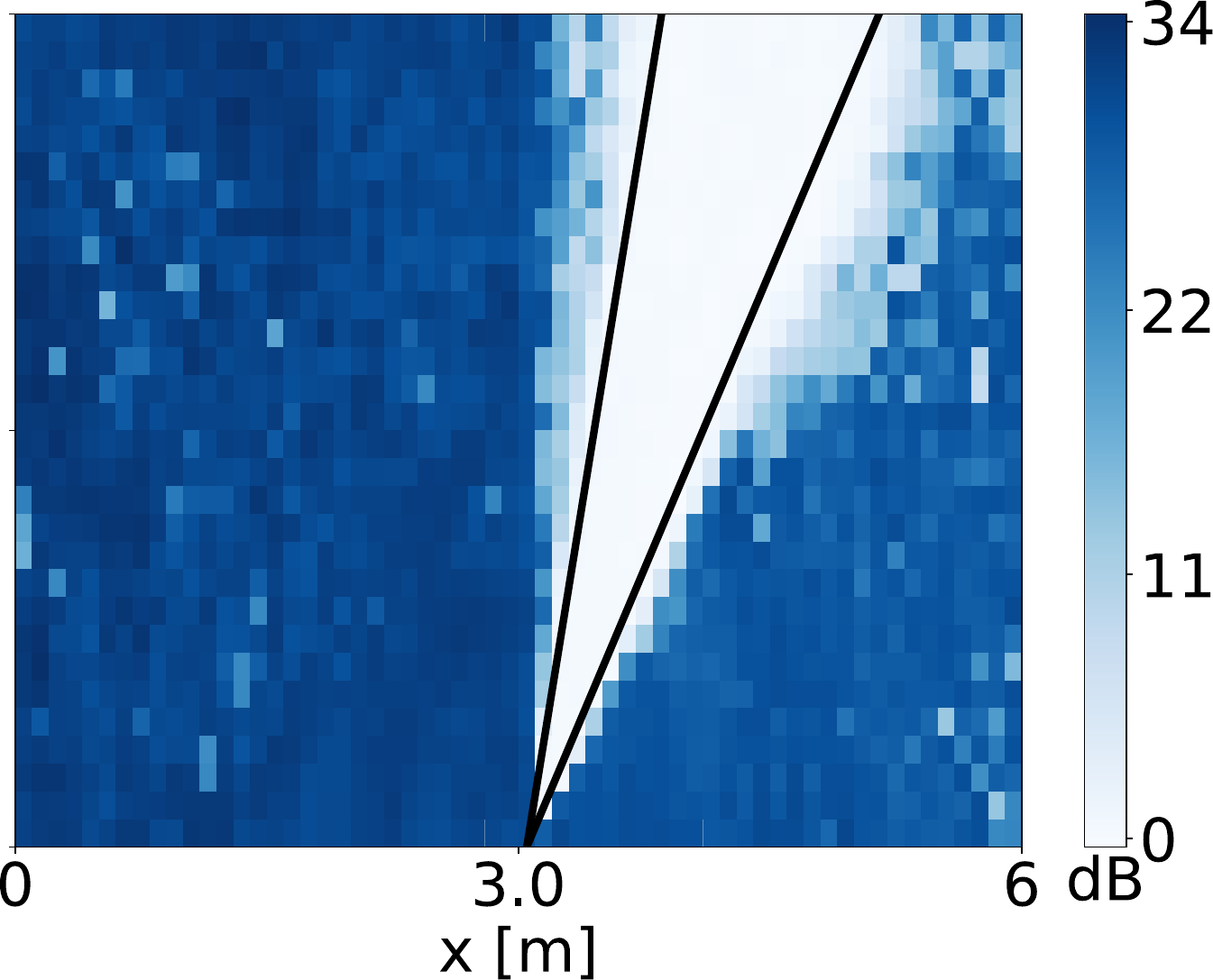}}}
  }
  \caption{PR heatmaps of the trained CRUSE$_{20}$ model when (a) no steering, (b) clockwise steering of $25^\circ$ and (c) $40^\circ$ is applied. (d) shows the baseline model CRUSE$_{\text{b-}20}$ that was trained to cover the ROI at a steered angle of $25^\circ$. The darker regions show stronger signal suppression. Black lines mark the original ROI, dashed orange lines mark the steered ROIs for completely linear steering, and solid red lines show the adapted ROI based on Equation~\ref{eq:boudnaries}. Due to front-back ambiguity, only one half of the room is shown.}
\label{fig:steering}
\end{figure*}

Let the central line of the learned ROI be at $\theta_1 = 90^{\circ}$ with respect to the microphone array. $\theta_1$ denotes the initial steering direction of the trained DNN, with the inter-microphone phase difference $\Delta\varphi_{\theta_1}=0$. As is illustrated in Figure~\ref{fig:setup}\,(b), we want to adapt $\theta_1$ by an angle $\gamma$, defining a new steering direction denoted as $\theta_2=\theta_1-\gamma$, with phase difference $\Delta\varphi_{\theta_2}$.

Any signal reaching the microphone array at $\theta_2$ must be adjusted with a phase shift $\Delta\psi$ so that it appears to the DNN to originate from $\theta_1$, i.e.,

\begin{equation}
     \Delta\varphi_{\theta_1}= \Delta\varphi_{\theta_2} + \Delta\psi
    \label{eq:psi}
\end{equation}
where $\Delta\psi =  2\pi f \frac{d}{c}(\cos(\theta_1) - \cos(\theta_2))$. This effectively simulates to the network that the signals in the steered area are located within the initially learned ROI.

The adapted spectrum $\tilde{\mathbf{Y}}_{m=2}$ is obtained by multiplying the STFT representation of the second microphone with a vector $\mathbf{a} \in \mathbb{C}^K$, with $K$ denoting the number of frequency bins, i.e.,

\begin{equation}
\begin{split}
    \mathbf{a}(k) = e^{j\Delta\psi} \phantom{l} \stackrel{\theta_1=90^{\circ}}{=} e^{-j2\pi f\frac{d}{c} \cos{\theta_2}}.
    \end{split}
\end{equation}

\begin{equation}
\tilde{\mathbf{Y}}_{m=2}(n,k) = \mathbf{a}(k)\mathbf{Y}_{m=2}(n,k).
\end{equation}

As is shown in Figure~\ref{fig:setup}\,(a), the initial ROI boundaries are defined by an angular width $\beta$. 
Then, $\Delta \varphi_{l}$ and $\Delta \varphi_{r}$ denote the phase differences present at the left and right ROI boundaries, defined as $ \Delta \varphi_{l,r} = 2\pi f \frac{d}{c} \cos (\theta_1 \pm \beta)$.

The corresponding phase differences $\Delta\varphi_{\phi_{l,r}}$ of the shifted ROI boundaries are now characterized by the angles $\phi_{l,r}$, i.e.,

\begin{equation}
    \Delta\varphi_{\phi_{l,r}} = 2\pi f \frac{d}{c} \cos(\phi_{l,r}).
\end{equation}

The phase differences at the initial ROI boundaries can then be expressed as $\Delta\varphi_{l,r}= \Delta\varphi_{\phi_{l,r}} + \Delta\psi$.
Using $\Delta\varphi_{l,r}$, Equation~\ref{eq:psi} and solving for $\phi_{l,r}$ results in 

\begin{align}
2\pi f \frac{d}{c} \cos(\theta_1 \pm \beta) &= 2\pi f \frac{d}{c} \cos(\phi_{l,r}) \notag\\ &+ 2\pi f \frac{d}{c}(\cos(\theta_1) - \cos(\theta_2)).\\
 \cos(\theta_1 \pm \beta) &= \cos(\phi_{l,r}) + 0 - \cos(\theta_1 -\gamma).\\
    \phi_{l,r}&=\arccos(\cos(\theta_1 \pm \beta) + \cos(\theta_1-\gamma)).
    \label{eq:boudnaries}
\end{align}

Two observations can  be made here: (i) steering the ROI is frequency independent, and (ii) the dependency of $\phi$ on the $\arccos$ function leads to a non-linear change in the shifted ROI boundaries, increasing the overall area. Figure~\ref{fig:doa}\,(b) illustrates this for $\beta=15^{\circ}$ and $\gamma=30^{\circ}$. 

\begin{table*}[!t]
     \centering
     \caption{DNSMOS and SI-SDR results (mean$\pm$std) for the test scenarios with different number of speakers and with/without noise. \textit{t} and \textit{k} represent target and interfering speech sources respectively. 1 denotes that a single source was present, whereas 23 represent randomly sampled 2 or 3 speakers.}
    \resizebox{\textwidth}{!}{
    \begin{tabular}{@{}c||l|c||cc|cc|cc|cc@{}}
    \toprule
     \textbf{Steering} & \multirow{2}{*}{\textbf{Method}} & \multirow{2}{*}{\textbf{Noise}}& \multicolumn{2}{c|}{\textbf{$\Delta$SIG}}  & \multicolumn{2}{c|}{\textbf{$\Delta$BAK}} & \multicolumn{2}{c|}{\textbf{$\Delta$OVRL}}  & \multicolumn{2}{c}{\textbf{$\Delta$SI-SDR}} \\
     \textbf{Angle} & && \multicolumn{1}{c}{\textbf{t1 k1}} & \multicolumn{1}{c|}{\textbf{t23 k23}} & \multicolumn{1}{c}{\textbf{t1 k1}} & \multicolumn{1}{c|}{\textbf{t23 k23}} & \multicolumn{1}{c}{\textbf{t1 k1}} & \multicolumn{1}{c|}{\textbf{t23 k23}} & \multicolumn{1}{c}{\textbf{t1 k1}} & \multicolumn{1}{c}{\textbf{t23 k23}}  \\ \midrule
     & CRUSE$_{\text{b-}20}$  &   \cmark    & \textbf{0.16} $\pm$ 0.60  & \textbf{0.99} $\pm$ 0.49 & 1.53 $\pm$ 0.53  & \textbf{1.65} $\pm$ 0.41 & \textbf{0.68} $\pm$ 0.34  & \textbf{0.73} $\pm$ 0.27 & 7.20 $\pm$ 2.52 & 2.98 $\pm$ 2.61 \\
        $25^\circ$       & CRUSE$_{20}$& \cmark & 0.12 $\pm$ 0.60  & 0.94 $\pm$ 0.52 & \textbf{1.60} $\pm$ 0.51  & 1.53 $\pm$ 0.52 & \textbf{0.68} $\pm$ 0.34 & 0.67 $\pm$ 0.32 & \textbf{7.42} $\pm$ 2.68 & \textbf{3.16} $\pm$ 2.65 \\ \midrule
     & CRUSE$_{\text{b-}40}$& \cmark & \textbf{0.38} $\pm$  0.77  & \textbf{1.08} $\pm$  0.53 & \textbf{1.70} $\pm$  0.63 & \textbf{1.70} $\pm$  0.52 & \textbf{0.78} $\pm$  0.47  & \textbf{0.80} $\pm$  0.36  & \textbf{7.66} $\pm$  2.20 & 3.18 $\pm$  3.33  \\
        $45^\circ$       & CRUSE$_{40}$& \cmark & 0.36 $\pm$  0.76  & 1.07 $\pm$  0.56 & 1.66 $\pm$  0.59 & 1.57 $\pm$  0.54 & 0.76 $\pm$  0.45 & 0.74 $\pm$  0.36 & 7.32 $\pm$  2.68 & \textbf{3.55} $\pm$  2.97  \\ \midrule \midrule
     & CRUSE$_{\text{b-}20}$ &  \xmark & \textbf{-0.05} $\pm$ 0.53  & 0.95 $\pm$ 0.60  & 1.27 $\pm$ 0.53 & \textbf{1.67} $\pm$ 0.49 & 0.50 $\pm$ 0.41  & \textbf{0.76} $\pm$ 0.36 &  \textbf{6.35} $\pm$ 2.03 & 0.93 $\pm$ 3.05 \\
        $25^\circ$       & CRUSE$_{20}$ & \xmark & -0.07 $\pm$ 0.51 & \textbf{0.98} $\pm$ 0.60 & \textbf{1.34} $\pm$ 0.55 & 1.62 $\pm$ 0.50 & \textbf{0.52} $\pm$ 0.42 & 0.74 $\pm$ 0.37 & 6.16 $\pm$ 2.58 & \textbf{1.04} $\pm$ 3.09  \\ \midrule
     & CRUSE$_{\text{b-}40}$ & \xmark & \textbf{0.08} $\pm$  0.57 & \textbf{1.06} $\pm$  0.52 & 1.35 $\pm$  0.59 & \textbf{1.74} $\pm$  0.39 & \textbf{0.62} $\pm$  0.43 & \textbf{0.81} $\pm$  0.32 & \textbf{7.30} $\pm$  2.21 & 1.47 $\pm$  3.33 \\
        $45^\circ$       & CRUSE$_{40}$ & \xmark &  0.04 $\pm$  0.57 & \textbf{1.06} $\pm$  0.51 & \textbf{1.37} $\pm$  0.56  & 1.59 $\pm$  0.50 & 0.60 $\pm$  0.42  & 0.76 $\pm$  0.31 & 6.72 $\pm$  2.50 & \textbf{1.56} $\pm$  2.92 \\ 
    
             \bottomrule
    \end{tabular}
    }
    \label{tab:spk_results}
    \end{table*}

\section{Experimental setup}
\subsection{Model architecture}
We used the CRUSE model \cite{cruse} for this work. It is a real-time capable U-Net architecture \cite{unet} operating in the STFT domain. As input, we concatenated the STFT representation of the stereo input along the channel dimension. Trained models generate a complex-valued single-channel separation mask, which was applied to the left channel of the input spectrum. Training used the AdamW optimizer \cite{loshchilov2018decoupled} with a 0.001 learning rate, $2e-05$ weight decay, and the SI-SDR \cite{sisdr} as a loss function. The STFT computation used a 20\,ms square-root Hann window (50\% overlap) and 320 NFFT points.

Overall, the model had approximately 640k parameters with 9.18\,GFLOPS in terms of computational complexity. A real-time factor (RTF) of 0.04 was obtained by computing the average processing time of 100 files of 10\,s length on a laptop with an 11th Gen. Intel(R) Core(TM) i7-1185G7 @ 3.00GHz.

\subsection{Training dataset}
A synthetic dataset was created using the \url{pyroomacoustics} package \cite{pra} to simulate the scenario. A two-microphone ULA was placed randomly in a shoebox room, with at least 2\,m distance to each wall. The room dimensions were uniformly sampled at [4.0\,m $\times$ 4.0\,m $\times$ 2.0\,m] to [8.0\,m $\times$ 8.0\,m $\times$ 4.0\,m] with T60 values randomly chosen from 0.25 to 0.7\,s to mimic an office-like scenario \cite{source_sep}. The microphone array and sources were assumed to be at the same height. Disjoint sets of speech and noise utterances from the DNS-Challenge dataset \cite{dubey2023icassp} at 16\,kHz were used for training. No sources were placed in the array’s mirrored ROI to avoid front-back ambiguity.

Each generated clip was 10\,s long and included one target and one interfering speaker, mixed at a signal-to-interference ratio (SIR) uniformly sampled between 0 and 10\,dB. Noise was added to the mix with SNR values that were sampled from a Gaussian $\mathcal{N}(7,3)$ distribution. The mix was level normalized using values from $\mathcal{N}(-28,10)$ dBFS, generating 55.6\,h of training data.

To understand how ROI size affects separation and steering performance, we trained two model versions, CRUSE$_{20}$ and CRUSE$_{40}$, with target area angles $\alpha=\{20^\circ, 40^\circ \}$. Both models used the same configuration but separate training sets.

\subsection{Test setup}
For CRUSE$_{20}$ and CRUSE$_{40}$ models, we generated separate test sets with the ROI steered $25^\circ$ and $45^\circ$ clockwise, respectively. Using speech from the 2020 Interspeech DNS-challenge \cite{reddy_interspeech} test set and noise from FSDnoisy18k \cite{fsdnoisy}, each test set included 4 scenarios varying the number of target and interfering speakers, with or without additional noise sources. Target sources were placed within the steered ROIs, and interfering sources remained in the initial ROI. This way it can be demonstrated that we can also mute the originally learned ROI using the proposed steering. Each test scenario included 100 clips.

As baseline methods two models termed CRUSE$_{\text{b-}20}$ and CRUSE$_{\text{b-}40}$ were trained to specifically cover the steered ROI without the proposed phase shift.

\subsection{Performance Evaluation}
We use power reduction (PR) heatmaps to visualize the effectiveness of our approach. Therefore, a single speech utterance is placed along the x and y direction in a room of size [6\,m $\times$ 6\,m $\times$ 3\,m], with a $T60 = 0.5\,s$. At each position, the PR metric \cite{patterson22_interspeech} is computed by
	\begin{equation}
		\text{PR}_{\text{dB}} := 10\log_{10} (||\mathbf{y}_{m=1}||^2 / ||\hat{\mathbf{t}}||^2),
	\end{equation}
\noindent where $\textbf{y}_{m=1}$ and $\hat{\textbf{t}}$ are the time domain representations of the mixture input signal at the reference microphone and the estimated target signal, respectively. Ideally, the PR should be strong outside the ROI and show no reduction inside. This setup enables us to create heatmaps to visualize the model's coverage of the ROI and its changes after steering.

In addition, we report the DNSMOS \cite{dnsmos} and SI-SDR \cite{sisdr} as evaluation metrics. 
DNSMOS is composed of sub-metrics evaluating the signal quality (SIG), background (BAK), and overall quality (OVRL). SI-SDR is the metric on which all models were optimized. The improvement ($\Delta$) compared to the input mixture signal is reported for all metrics.

\begin{figure}[]
\centering
\includegraphics[width=\linewidth]{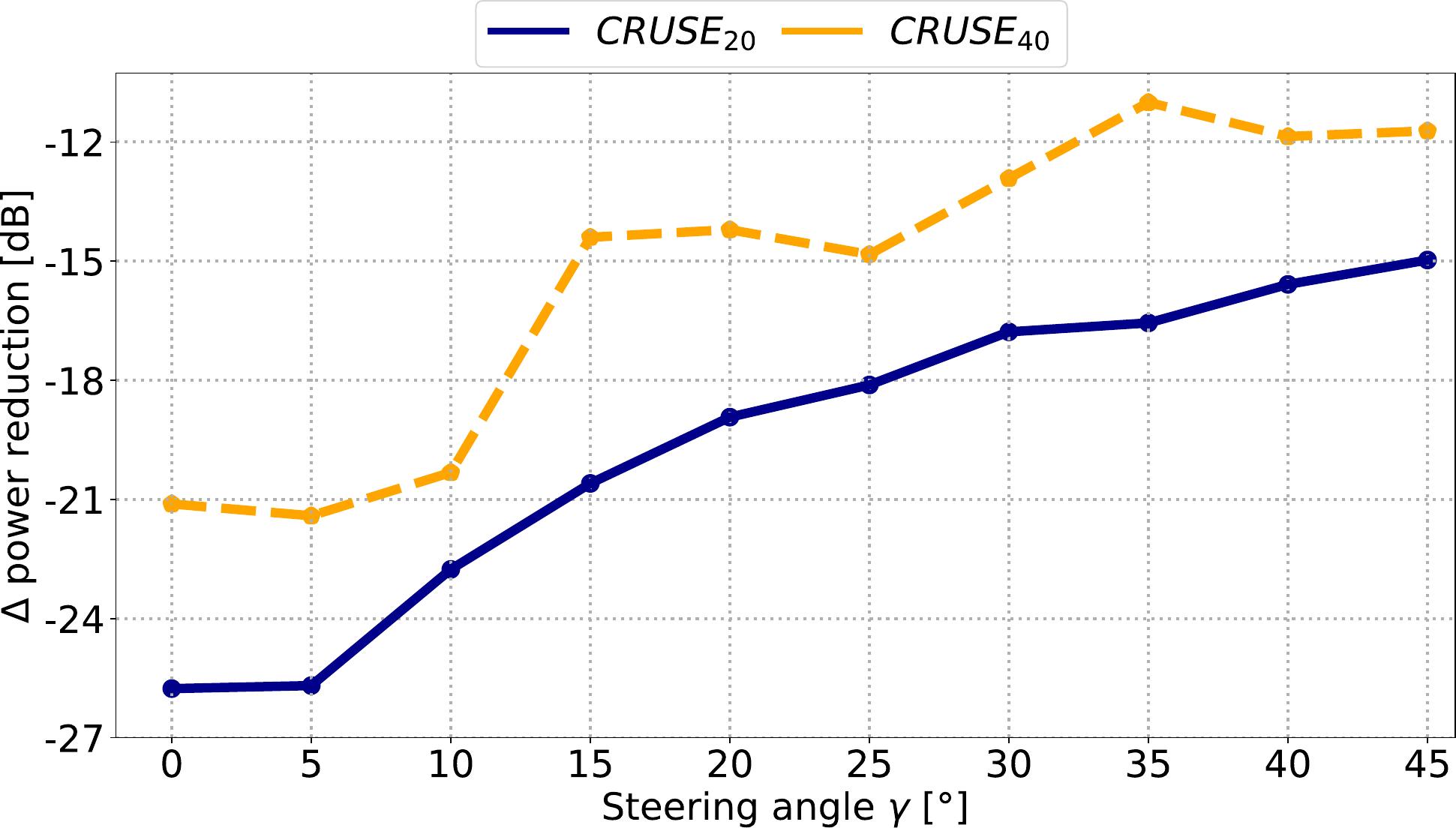}
\caption{PR performance of CRUSE$_{\{20,40\}}$ models at different steering angles. $\Delta$PR is computed as the difference of mean PR in the ROI and the area outside.}
\label{fig:steer_vs_pr}
\end{figure}

\section{Results and Discussion}
We first investigated the \textit{Neural Steering} performance with PR heatmaps in Figure \ref{fig:steering}. The heatmaps were generated using the CRUSE$_{20}$ model for steering angles of $0^\circ$ (i.e., no steering), $25^\circ$ and $40^\circ$. In addition, the heatmap of the corresponding baseline model is shown. 
This analysis demonstrates that \textit{Neural Steering} can successfully steer the ROI in new directions.

Second, we investigated the performance of \textit{Neural Steering} for different steering angles in Figure \ref{fig:steer_vs_pr}.  The figure shows the difference of mean PR of all positions within the ROI and the area outside against the steering angles for the CRUSE$_{\{20,40\}}$ models. The steering was evaluated in five degrees steps until 45$^\circ$. We observe that the PR is the strongest for the initial learned ROI at $\gamma=0$ and decreases with increasing steering values. This demonstrates that with increasing steering angles the non-linear mapping of the ROI also leads to less suppression close to the ROI boundaries. This effect can also be found in Figure \ref{fig:steering}, where the brighter area indicating low suppression increases with an increasing steering angle. 
Notably, the precision of the steered ROI decreases as the sources move further from the microphone array and the steering angle increases.

Lastly, we investigated the performance of \textit{Neural Steering} in terms of DNSMOS and SI-SDR compared to the baseline approaches in Table \ref{tab:spk_results}. For this analysis, we investigated the CRUSE$_{20}$, CRUSE$_{40}$  models' performance for steering angles of 25$^\circ$ and 45$^\circ$. 
The baseline models are trained with their respective ROI at these angles. 

In general, all approaches show better improvements in `\textit{with noise}' scenarios compared to `\textit{without noise}', indicating their capability to suppress noise. Overall, the proposed \textit{Neural Steering} approach on CRUSE models shows comparable results to baselines in all scenarios. This demonstrates that our approach can steer to new target directions at negligible cost, reinforcing its practicality without the need for retraining the DNN for each direction.


\section{Conclusion}
This work proposed \textit{Neural Steering}, an inference-adaptive steering method for real-time audio source separation. This is achieved by applying a phase shift to the captured audio, which steers a trained DNN to a new target area, including the desired speech utterances. In the future, we intend to combine this approach with a direction-of-arrival estimator to track moving target speakers such that steering can be applied dynamically over time. 
\newpage
\balance
\bibliographystyle{IEEEtran}
\bibliography{refs}

\end{document}

%% file: setup.tikz
\tikzset{every picture/.style={line width=0.75pt}} 

\begin{tikzpicture}[x=0.75pt,y=0.75pt,yscale=-1,xscale=1]

\draw  [color={rgb, 255:red, 143; green, 203; blue, 75 }  ,draw opacity=1 ][fill={rgb, 255:red, 184; green, 233; blue, 134 }  ,fill opacity=0.28 ][dash pattern={on 4.5pt off 4.5pt}] (163.87,170.36) -- (196.92,234.89) -- (131,234.98) -- cycle ;
\draw  [fill={rgb, 255:red, 0; green, 0; blue, 0 }  ,fill opacity=1 ] (155.83,170.4) .. controls (155.83,169.02) and (156.95,167.9) .. (158.33,167.9) .. controls (159.71,167.9) and (160.83,169.02) .. (160.83,170.4) .. controls (160.83,171.78) and (159.71,172.9) .. (158.33,172.9) .. controls (156.95,172.9) and (155.83,171.78) .. (155.83,170.4) -- cycle ;
\draw  [fill={rgb, 255:red, 0; green, 0; blue, 0 }  ,fill opacity=1 ] (168.5,170.23) .. controls (168.5,168.85) and (169.62,167.73) .. (171,167.73) .. controls (172.38,167.73) and (173.5,168.85) .. (173.5,170.23) .. controls (173.5,171.61) and (172.38,172.73) .. (171,172.73) .. controls (169.62,172.73) and (168.5,171.61) .. (168.5,170.23) -- cycle ;
\draw  [color={rgb, 255:red, 65; green, 117; blue, 5 }  ,draw opacity=1 ][fill={rgb, 255:red, 184; green, 233; blue, 134 }  ,fill opacity=0.6 ] (164.09,170.37) -- (130.78,105.86) -- (197.15,105.74) -- cycle ;
\draw  [dash pattern={on 0.84pt off 2.51pt}]  (164.88,106.29) -- (164.88,136.6) ;
\draw  [draw opacity=0] (151.88,147.09) .. controls (154.75,142.2) and (159.64,139.01) .. (165.15,139.07) .. controls (169.95,139.13) and (174.22,141.65) .. (177.06,145.58) -- (164.93,157.11) -- cycle ; \draw   (151.88,147.09) .. controls (154.75,142.2) and (159.64,139.01) .. (165.15,139.07) .. controls (169.95,139.13) and (174.22,141.65) .. (177.06,145.58) ;  
\draw   (98.48,105.29) -- (228.61,105.29) -- (228.61,235.42) -- (98.48,235.42) -- cycle ;
\draw    (137.21,146.42) -- (161.51,131.47) ;
\draw [shift={(163.21,130.42)}, rotate = 148.39] [fill={rgb, 255:red, 0; green, 0; blue, 0 }  ][line width=0.08]  [draw opacity=0] (8,-3) -- (0,0) -- (8,3) -- cycle    ;

\draw (147.01,115.52) node [anchor=north west][inner sep=0.75pt]  [font=\scriptsize,color={rgb, 255:red, 95; green, 124; blue, 65 }  ,opacity=1 ] [align=left] {\textbf{X}};
\draw (189.01,130.86) node [anchor=north west][inner sep=0.75pt]  [font=\scriptsize,color={rgb, 255:red, 208; green, 2; blue, 27 }  ,opacity=1 ] [align=left] {\textbf{X}};
\draw (100.68,113.72) node [anchor=north west][inner sep=0.75pt]  [font=\scriptsize]  {$\theta _{1} +\beta $};
\draw (194.01,115.06) node [anchor=north west][inner sep=0.75pt]  [font=\scriptsize]  {$\theta _{1} -\beta $};
\draw (160.34,147.06) node [anchor=north west][inner sep=0.75pt]  [font=\scriptsize]  {$\alpha $};
\draw (126.34,175.37) node [anchor=north west][inner sep=0.75pt]  [font=\scriptsize] [align=left] {Microphone array};
\draw (101.68,157.72) node [anchor=north west][inner sep=0.75pt]  [font=\scriptsize]  {$\theta _{1} =90^{\circ }$};
\draw (100.34,136.37) node [anchor=north west][inner sep=0.75pt]  [font=\scriptsize] [align=left] {Steering \\direction};
\draw (180.34,191.52) node [anchor=north west][inner sep=0.75pt]  [font=\scriptsize] [align=left] {\textbf{X: Speech}};

\end{tikzpicture}

%% file: setup_steered.tikz
\tikzset{every picture/.style={line width=0.75pt}} 

\begin{tikzpicture}[x=0.75pt,y=0.75pt,yscale=-1,xscale=1]

\draw  [color={rgb, 255:red, 143; green, 203; blue, 75 }  ,draw opacity=1 ][fill={rgb, 255:red, 184; green, 233; blue, 134 }  ,fill opacity=0.28 ][dash pattern={on 4.5pt off 4.5pt}] (228.44,201.35) -- (164.59,169.92) -- (172.04,235.19) -- (228.61,235.42) -- cycle ;
\draw  [color={rgb, 255:red, 65; green, 117; blue, 5 }  ,draw opacity=1 ][fill={rgb, 255:red, 184; green, 233; blue, 134 }  ,fill opacity=0.6 ] (222.55,105.3) -- (228.61,105.29) -- (228.84,140.39) -- (164.59,169.92) -- (172.04,105.59) -- (181.15,105.3) -- cycle ;
\draw  [fill={rgb, 255:red, 0; green, 0; blue, 0 }  ,fill opacity=1 ] (155.83,170.4) .. controls (155.83,169.02) and (156.95,167.9) .. (158.33,167.9) .. controls (159.71,167.9) and (160.83,169.02) .. (160.83,170.4) .. controls (160.83,171.78) and (159.71,172.9) .. (158.33,172.9) .. controls (156.95,172.9) and (155.83,171.78) .. (155.83,170.4) -- cycle ;
\draw  [fill={rgb, 255:red, 0; green, 0; blue, 0 }  ,fill opacity=1 ] (168.5,170.23) .. controls (168.5,168.85) and (169.62,167.73) .. (171,167.73) .. controls (172.38,167.73) and (173.5,168.85) .. (173.5,170.23) .. controls (173.5,171.61) and (172.38,172.73) .. (171,172.73) .. controls (169.62,172.73) and (168.5,171.61) .. (168.5,170.23) -- cycle ;
\draw   (98.48,105.29) -- (228.61,105.29) -- (228.61,235.42) -- (98.48,235.42) -- cycle ;
\draw  [dash pattern={on 0.84pt off 2.51pt}]  (164.59,169.92) -- (206.4,105.47) ;
\draw  [dash pattern={on 0.84pt off 2.51pt}]  (165.26,105.25) -- (164.59,169.92) ;
\draw    (165.26,122.6) .. controls (176.17,104.17) and (188.49,107.85) .. (190.37,127.94) ;
\draw [shift={(190.51,129.85)}, rotate = 265.37] [fill={rgb, 255:red, 0; green, 0; blue, 0 }  ][line width=0.08]  [draw opacity=0] (8,-3) -- (0,0) -- (8,3) -- cycle    ;

\draw (147.01,115.52) node [anchor=north west][inner sep=0.75pt]  [font=\scriptsize,color={rgb, 255:red, 208; green, 2; blue, 27 }  ,opacity=1 ] [align=left] {\textbf{X}};
\draw (189.01,130.86) node [anchor=north west][inner sep=0.75pt]  [font=\scriptsize,color={rgb, 255:red, 95; green, 124; blue, 65 }  ,opacity=1 ] [align=left] {\textbf{X}};
\draw (99.98,151.62) node [anchor=north west][inner sep=0.75pt]  [font=\scriptsize]  {$\theta _{2} =90^{\circ } -\gamma $};
\draw (126.34,175.37) node [anchor=north west][inner sep=0.75pt]  [font=\scriptsize] [align=left] {Microphone array};
\draw (99.94,192.32) node [anchor=north west][inner sep=0.75pt]  [font=\scriptsize] [align=left] {\textbf{X: Speech}};
\draw (99.85,129.1) node [anchor=north west][inner sep=0.75pt]  [font=\scriptsize] [align=left] {Steering\\direction};
\draw (172.79,113.74) node [anchor=north west][inner sep=0.75pt]  [font=\scriptsize]  {$\gamma $};

\end{tikzpicture}

%% file: doa_2.tikz
\tikzset{every picture/.style={line width=0.75pt}} 

\begin{tikzpicture}[x=0.75pt,y=0.75pt,yscale=-1,xscale=1]

\draw  [color={rgb, 255:red, 0; green, 0; blue, 0 }  ,draw opacity=0.5 ][dash pattern={on 3.75pt off 3pt on 7.5pt off 1.5pt}] (153.23,98.41) .. controls (148.07,95.8) and (143.15,93.32) .. (141.92,94.64) .. controls (140.68,95.96) and (143.46,100.71) .. (146.38,105.7) .. controls (149.31,110.69) and (152.09,115.44) .. (150.85,116.76) .. controls (149.62,118.08) and (144.7,115.6) .. (139.54,112.99) .. controls (134.38,110.38) and (129.46,107.9) .. (128.22,109.21) .. controls (126.98,110.53) and (129.76,115.29) .. (132.69,120.28) .. controls (135.62,125.26) and (138.4,130.02) .. (137.16,131.34) .. controls (135.92,132.66) and (131,130.17) .. (125.84,127.56) .. controls (120.68,124.95) and (115.76,122.47) .. (114.52,123.79) .. controls (113.28,125.11) and (116.07,129.86) .. (118.99,134.85) .. controls (121.92,139.84) and (124.7,144.59) .. (123.46,145.91) .. controls (122.22,147.23) and (117.31,144.75) .. (112.14,142.14) .. controls (106.98,139.53) and (102.07,137.04) .. (100.83,138.36) .. controls (99.85,139.4) and (101.38,142.59) .. (103.51,146.34) ;
\draw   (150.54,97.08) -- (201.08,148) -- (100,148) -- cycle ;
\draw  [fill={rgb, 255:red, 0; green, 0; blue, 0 }  ,fill opacity=1 ] (94.73,148) .. controls (94.73,145.09) and (97.09,142.73) .. (100,142.73) .. controls (102.91,142.73) and (105.27,145.09) .. (105.27,148) .. controls (105.27,150.91) and (102.91,153.27) .. (100,153.27) .. controls (97.09,153.27) and (94.73,150.91) .. (94.73,148) -- cycle ;
\draw  [draw opacity=0] (164.56,110.35) .. controls (161.04,112.9) and (156.07,114.51) .. (150.55,114.56) .. controls (145.5,114.61) and (140.88,113.34) .. (137.4,111.22) -- (150.43,101.32) -- cycle ; \draw   (164.56,110.35) .. controls (161.04,112.9) and (156.07,114.51) .. (150.55,114.56) .. controls (145.5,114.61) and (140.88,113.34) .. (137.4,111.22) ;  
\draw  [fill={rgb, 255:red, 0; green, 0; blue, 0 }  ,fill opacity=1 ] (149.68,106.82) .. controls (149.68,106.3) and (150.1,105.89) .. (150.61,105.89) .. controls (151.13,105.89) and (151.55,106.3) .. (151.55,106.82) .. controls (151.55,107.33) and (151.13,107.75) .. (150.61,107.75) .. controls (150.1,107.75) and (149.68,107.33) .. (149.68,106.82) -- cycle ;
\draw  [draw opacity=0] (119.7,128.05) .. controls (124.29,126.98) and (129.39,129.27) .. (131.97,133.94) .. controls (134.73,138.94) and (133.64,144.94) .. (129.64,148.13) -- (122.73,139.03) -- cycle ; \draw   (119.7,128.05) .. controls (124.29,126.98) and (129.39,129.27) .. (131.97,133.94) .. controls (134.73,138.94) and (133.64,144.94) .. (129.64,148.13) ;  
\draw   (150.66,96.98) .. controls (155.59,100.01) and (160.28,102.89) .. (161.63,101.67) .. controls (162.97,100.46) and (160.59,95.49) .. (158.09,90.28) .. controls (155.58,85.07) and (153.2,80.1) .. (154.55,78.89) .. controls (155.89,77.68) and (160.59,80.56) .. (165.51,83.59) .. controls (170.44,86.62) and (175.14,89.5) .. (176.48,88.28) .. controls (177.83,87.07) and (175.45,82.1) .. (172.94,76.89) .. controls (170.44,71.68) and (168.06,66.71) .. (169.4,65.5) .. controls (170.75,64.29) and (175.44,67.17) .. (180.37,70.2) .. controls (185.3,73.23) and (189.99,76.11) .. (191.34,74.89) .. controls (192.68,73.68) and (190.3,68.71) .. (187.8,63.5) .. controls (185.29,58.29) and (182.92,53.32) .. (184.26,52.11) .. controls (184.89,51.54) and (186.26,51.87) .. (188.05,52.7) ;
\draw  [fill={rgb, 255:red, 0; green, 0; blue, 0 }  ,fill opacity=1 ] (195.81,148) .. controls (195.81,145.09) and (198.17,142.73) .. (201.08,142.73) .. controls (204,142.73) and (206.36,145.09) .. (206.36,148) .. controls (206.36,150.91) and (204,153.27) .. (201.08,153.27) .. controls (198.17,153.27) and (195.81,150.91) .. (195.81,148) -- cycle ;
\draw   (175.52,121.83) .. controls (180.44,124.86) and (185.14,127.74) .. (186.48,126.53) .. controls (187.83,125.32) and (185.45,120.35) .. (182.94,115.14) .. controls (180.44,109.92) and (178.06,104.96) .. (179.41,103.74) .. controls (180.75,102.53) and (185.45,105.41) .. (190.37,108.44) .. controls (195.3,111.47) and (199.99,114.35) .. (201.34,113.14) .. controls (202.68,111.93) and (200.3,106.96) .. (197.8,101.75) .. controls (195.3,96.53) and (192.92,91.57) .. (194.26,90.35) .. controls (195.61,89.14) and (200.3,92.02) .. (205.23,95.05) .. controls (210.15,98.08) and (214.85,100.96) .. (216.19,99.75) .. controls (217.54,98.54) and (215.16,93.57) .. (212.66,88.36) .. controls (210.15,83.14) and (207.77,78.18) .. (209.12,76.96) .. controls (209.75,76.39) and (211.12,76.73) .. (212.91,77.56) ;
\draw   (201.23,147.26) .. controls (206.16,150.29) and (210.85,153.17) .. (212.2,151.96) .. controls (213.54,150.75) and (211.16,145.78) .. (208.66,140.57) .. controls (206.15,135.35) and (203.78,130.38) .. (205.12,129.17) .. controls (206.46,127.96) and (211.16,130.84) .. (216.09,133.87) .. controls (221.01,136.9) and (225.71,139.78) .. (227.05,138.57) .. controls (228.4,137.36) and (226.02,132.39) .. (223.51,127.18) .. controls (221.01,121.96) and (218.63,116.99) .. (219.98,115.78) .. controls (221.32,114.57) and (226.02,117.45) .. (230.94,120.48) .. controls (235.87,123.51) and (240.56,126.39) .. (241.91,125.18) .. controls (243.25,123.97) and (240.87,119) .. (238.37,113.79) .. controls (235.87,108.57) and (233.49,103.6) .. (234.83,102.39) .. controls (235.46,101.82) and (236.83,102.16) .. (238.62,102.99) ;

\draw (121.7,131.45) node [anchor=north west][inner sep=0.75pt]  [font=\footnotesize]  {$\mathnormal{{\textstyle \theta }}$};
\draw (145.71,149.98) node [anchor=north west][inner sep=0.75pt]    {$d$};
\draw (118.29,106.48) node [anchor=north west][inner sep=0.75pt]    {$l$};

\end{tikzpicture}

%% file: arccos.tikz
\begin{tikzpicture}
  \begin{axis}
    [domain = -0.1:1, 
    samples = 100,
    axis lines=center,
    axis line style={black, thick, line width=.06cm},
    tick style={line width=.06cm, color=black, line cap=round},
    font=\Large,
    extra x ticks = {0},
    xmin=-0.02,
    xmax=1.1,
    ymin=-0.1,
    ymax=100,
    xtick={0,0.5,1},
    ytick={0,60, 90},
    yticklabels={0, $\frac{1}{3}\pi$,$\frac{1}{2}\pi$},
    legend style={at={(0.1,0.1)},anchor=west},
    extra y ticks={40.64,76.04},
    extra y tick labels={$\phi_r$, $\phi_l$},
    ]
    \addplot[color = black,thick, line width=.06cm] {acos(x)};
    \draw [thick]   (60,58) -- (60,62) ;
    \draw [thick]    (58,60) -- (62,60) ;

    \draw [color={rgb, 255:red, 128; green, 128; blue, 128 }  ,draw opacity=1 ][fill={rgb, 255:red, 128; green, 128; blue, 128 }  ,fill opacity=1 , thick]   (10,90) -- (60,90) ;
    \draw [color={rgb, 255:red, 128; green, 128; blue, 128 }  ,draw opacity=1 ][fill={rgb, 255:red, 155; green, 155; blue, 155 }  ,fill opacity=1 , thick]  (60,90) -- (60,65) ;
    \draw [fill={rgb, 255:red, 128; green, 128; blue, 128 }  ,fill opacity=1 ][line width=0.08]  [draw opacity=0] (58,65) -- (60,60) -- (62,65) -- cycle    ;

    \draw (12,100) node [anchor=north west][inner sep=0.75pt]    {$\cos\left( 90^{\circ } -\gamma \right)$};

    \draw [thick]   (34.12,72) -- (34.12,80) ;
    \draw [thick]   (85.88,37) -- (85.88,45) ;
    \draw  [dash pattern={on 0.84pt off 2.51pt}, thick]  (34.12,76.04) -- (10,76.04) ;
    \draw  [dash pattern={on 0.84pt off 2.51pt}, thick]  (85.88,40.64) -- (10,40.64) ;

    \addlegendentry{$\arccos(x)$}

    \draw [color={rgb, 255:red, 128; green, 128; blue, 128 }  ,draw opacity=1 ][fill={rgb, 255:red, 128; green, 128; blue, 128 }  ,fill opacity=1 , thick]  (60,60) -- (80.88, 60) ;
    \draw [color={rgb, 255:red, 128; green, 128; blue, 128 }  ,draw opacity=1 ][fill={rgb, 255:red, 128; green, 128; blue, 128 }  ,fill opacity=1 , thick]  (60,60) -- (39.12, 60) ;
    \draw [fill={rgb, 255:red, 128; green, 128; blue, 128 }  ,fill opacity=1 ][line width=0.08]  [draw opacity=0] (80.88,62) -- (80.88,58) -- (85.88,60) -- cycle    ;
    \draw [fill={rgb, 255:red, 128; green, 128; blue, 128 }  ,fill opacity=1 ][line width=0.08]  [draw opacity=0] (34.12,60) -- (39.12,58) -- (39.12,62) -- cycle    ;
    \draw [color={rgb, 255:red, 128; green, 128; blue, 128 }  ,draw opacity=1 ][fill={rgb, 255:red, 155; green, 155; blue, 155 }  ,fill opacity=1 , thick]  (85.88,58) -- (85.88,62) ;
    \draw [color={rgb, 255:red, 128; green, 128; blue, 128 }  ,draw opacity=1 ][fill={rgb, 255:red, 155; green, 155; blue, 155 }  ,fill opacity=1 , thick]  (34.12,58) -- (34.12,62) ;

    \draw (65,70) node [anchor=north west][inner sep=0.75pt]    {$+ \cos\left( 90^{\circ } \pm \beta \right)$};
    
  \end{axis}

\end{tikzpicture}